\documentclass[12pt]{iopart}

\usepackage{iopams}
\usepackage{graphicx}
\expandafter\let\csname equation*\endcsname\relax
\expandafter\let\csname endequation*\endcsname\relax
\usepackage{amsmath}
\usepackage{bm}

\usepackage[normalem]{ulem}
\usepackage{xcolor}

\begin{document}

\title{ Infinitesimal asphericity changes the universality of the
 jamming transition}
 \author{Harukuni Ikeda$^1$, Carolina Brito$^2$, Matthieu Wyart$^3$}

\address{
$^1$Laboratoire de Physique de l'Ecole Normale Sup\'erieure, ENS, Universit\'e PSL, CNRS, Sorbonne Universit\'e,
Universit\'e de Paris,
Paris, France \\
$^2$Instituto de F\'isica, UFRGS, 91501-970, Porto Alegre, Brazil\\
$^3$Institute of Physics, EPFL, CH-1015 Lausanne, Switzerland
}
\ead{harukuni.ikeda@lpt.ens.fr}

\vspace{10pt}
\begin{indented}
\item[]March 2019
\end{indented}

\begin{abstract} 
The jamming transition of non-spherical particles is fundamentally
different from the spherical case. Non-spherical
particles are hypostatic at their jamming points, while isostaticity is
ensured in the case of the jamming of spherical particles.  This
structural difference implies that the presence of asphericity affects
the critical exponents related to the contact number and the vibrational
density of states. Moreover, while the force and gap distributions of
isostatic jamming present power-law behaviors, even an infinitesimal
asphericity is enough to smooth out these singularities. In a recent
work [PNAS 115(46), 11736], we have used a combination of marginal
stability arguments and the replica method to explain these
observations. We argued that systems with internal degrees of freedom,
like the rotations in ellipsoids, or the variation of the radii in the
case of the \textit{breathing} particles fall in the same universality
class. In this paper, we review comprehensively the results about the
jamming with internal degrees of freedom in addition to the
translational degrees of freedom.  We use a variational argument to
derive the critical exponents of the contact number, shear modulus, and
the characteristic frequencies of the density of states. Moreover, we
present additional numerical data supporting the theoretical results,
which were not shown in the previous work.
\end{abstract}

%
%
%
%
%

\newcommand{\diff}[2]{\frac{d#1}{d#2}}
\newcommand{\pdiff}[2]{\frac{\partial #1}{\partial #2}}
\newcommand{\fdiff}[2]{\frac{\delta #1}{\delta #2}}
\newcommand{\new}{\nonumber\\}
\newcommand{\bx}{\bm{x}}
\newcommand{\bu}{\bm{u}}
\newcommand{\hr}{\hat{r}}
\newcommand{\hf}{\hat{f}}
\newcommand{\hpi}{\hat{\pi}}
\newcommand{\bX}{\bm{X}}
\newcommand{\tk}{\tilde{k}}
\newcommand{\abs}[1]{\left|#1\right|}
\newcommand{\ave}[1]{\left\langle #1\right\rangle}
\newcommand{\M}{\mathcal{M}}
\newcommand{\Y}{\mathcal{Y}}
\newcommand{\A}{\mathcal{A}}
\newcommand{\Z}{\mathcal{Z}}
\newcommand{\G}{\mathcal{G}}
\newcommand{\FF}{\mathcal{F}}
\newcommand{\he}{\mathcal{H}}
\newcommand{\im}{{\rm Im}}

\section{Introduction}

The jamming transition of non-spherical particles is qualitatively
different from that of spherical
particles~\cite{van2009,torquato2010}. Several experimental and
numerical investigations uncover that (i) systems consisting of
non-spherical particles are not isostatic at their jamming transition
point~\cite{donev2004}, while systems of spherical particles are
isostatic~\cite{bernal1960}, (ii) the pair correlation of non-spherical
particles does not exhibit the power law singularity at the jamming
transition point~\cite{brito2018universality}, while that of spherical
particles does~\cite{donev2005}, and (iii) the critical exponents of
non-spherical particles are different from those of spherical
particles~\cite{mailman2009,schreck2012}.

The theoretical understanding of the jamming of non-spherical particles
is challenging because particles do not hold rotational symmetry. In the
previous work~\cite{brito2018universality,ikeda2019mean}, we proposed a
way to bypass this difficulty by considering the mapping from
non-spherical particles to the breathing particles (BP), defined as a
system of spherical particles for which their radii are allowed to
fluctuate~\cite{BP2018}. An advantage of the BP particles is that the
model holds the rotational symmetry, and thus, one can apply the same
technique developed for the spherical particle without any
difficulty. Using the BP, we theoretically and numerically confirmed
that the gap and force distributions of non-spherical particles are
regular and finite even at the jamming transition point, while those
quantities exhibit the power law in the case of spherical
particles. Furthermore, we showed that the critical exponents of several
physical quantities, such as the contact number, shear modulus, and
characteristic frequencies of the density of states, have different
values from those of spherical particles. This confirms that the jamming
of non-spherical particles belongs to a different universality class
from that of spherical particles.

This paper is a longer version of our previous
work~\cite{brito2018universality}. We shall give a more straight forward
derivation of the scaling functions without mapping to the BP particles,
and additional numerical data supporting the theoretical results. The
organization of the remaining paper is as follows.  In
Sec.~\ref{152654_8Jul19}, we develop a variational argument for
non-spherical particles.  In Sec.~\ref{152720_8Jul19}, we discuss the
connection between nonspherial particles and BP. In
Sec.~\ref{152733_8Jul19}, we discuss the universal form of the gap and
force distributions near the isostatic point.  In
Sec.~\ref{163348_3Jul19}, we discuss the scaling behavior of the density
of states of the BP and show that the characteristic frequencies exhibit
the same scaling as non-spherical particles. In
Sec.~\ref{152804_8Jul19}, we summarize and conclude the work.

\section{Variational argument}
\label{152654_8Jul19} Here we derive the scaling functions of
non-spherical particles for a small asphericity by using the variational
argument~\cite{wyart2005effects,yan2016}. In the previous
work~\cite{brito2018universality}, we performed this calculation by
mapping the Gay-Berne potential, which is a model for ellipsoids, to the
breathing particles (BP), which is the model consisting of spherical
particles where the radii of particles can vary
continuously~\cite{BP2018}. In this paper, instead, we present a more
direct derivation of the scaling functions of non-spherical particles
without using the mapping to the BP model.

\subsection{Interaction potential}
  For concreteness, we consider the following
interaction potential:
\begin{align}
V_N &= \sum_{i<j}v(h_{ij}),\label{235848_23Jan19}
\end{align}
where $h_{ij}$ denotes the minimal distance between the $i$-th and
$j$-th particles, and $v(h)$ denotes a purely repulsive and finite
ranged potential, such as the harmonic potential $v(h)=h^2\theta(-h)/2$,
where $\theta(x)$ denotes the Heaviside function.

\subsection{Perturbation around spherical particles}
We first derive a convenient expression of interaction potential for a
small asphericity.  Non-spherical particle have rotational degrees of
freedom in addition to the translational degrees of freedom. We assign
the vectors $\bx_i$ and a unit vector $\bm{u}_i$ to express the position
and direction of the $i$-th particle, respectively. The radius
$\sigma_i$ of a non-spherical particle along the direction $\hat{r}$
varies depending on both $\bm{u}_i$ and $\hat{r}$. We shall assume that
there is a small parameter $\Delta$ representing the deviation from a
sphere. We expand the radius using $\Delta$ as
\begin{align}
 \sigma_i(\hat{r},\bm{u}_i) = \sigma^{(0)} + f(\hat{r},\bm{u}_i)\Delta + O(\Delta^2),
\end{align}
where $\sigma^{(0)}$ represents the radius of the reference sphere,
and $f(\hat{r},\bm{u})$ represents the coefficient of the first order
term.  Following the similar procedure, one can expand the gap function,
which is the minimal distance between the $i$-th and $j$-th particles.
The first order correction of $\Delta$ comes from the change of radii
of the $i$-th and $j$-th particles along the direction
$\hr_{ij}=(\bx_i-\bx_j)/\abs{\bx_i-\bx_j}$, namely,
\begin{align}
 h_{ij}(\Delta) -h_{ij}(0)
 &=
 -\Delta \left[f(\hat{r}_{ij},\bm{u}_i) + f(-\hat{r}_{ij},\bm{u}_j)\right]+ O(\Delta^2),
\end{align}
where $h_{ij}(0)$ is the gap function of the reference spherical
particles:
\begin{align}
h_{ij}(0) = r_{ij} -\sigma_i^{(0)}-\sigma_j^{(0)}.
\end{align}
Substituting this into Eq.~(\ref{235848_23Jan19}) and expanding
by $\Delta$, we have
\begin{align}
V_N &= U_N + Q_N,\label{165606_24Jan19}
\end{align}
where
\begin{align}
 U_N &= \sum_{i<j}\left[v(h_{ij}^{(0)}) +  w_{ij}\Delta^2\right],\new
 Q_N 
 &= \sum_i g_i(\bu_i).\label{165758_24Jan19}
\end{align}
Here $ w_{ij} \Delta^2$ denotes the $O(\Delta^2)$ term of the
interaction potential, and we have introduced the auxiliary function as
\begin{align}
 g_i(\bu_i) &= -\Delta\sum_{j\neq i}v'(h_{ij}(0))
 \left[f(\hr_{ij},\bu_i)+f(-\hr_{ij},\bu_i)\right].\label{165611_24Jan19}
\end{align}
From Eqs.~(\ref{165758_24Jan19}) and (\ref{165611_24Jan19}), 
one can show that
\begin{align}
 Q_N \sim g_i \sim p\Delta,\label{184142_24Jan19}
\end{align}
where $p\sim - v'(h)$ denotes the pressure.

To understand the physical meaning of $\Delta$, it is convenient to
clarify the relation between $\Delta$ and the sphericity $\A$, which
represents how far is the shape of a particle from a perfect sphere;
$\A=1$ for a perfect sphere and $\A>1$ otherwise.  By definition, $\A$
takes a minimal value at $\Delta=0$, which allows us to expand it as
$\A(\Delta) = 1 + \frac{1}{2}\A''(0)\Delta^2 + O(\Delta^3)$, leading to
\begin{align}
\Delta \sim (\A-1)^{1/2}. 
\end{align}
This is a useful relation to compare with numerical and experimental
results.

\subsection{Variational argument for spherical particles}

In this work, we derive the scaling behaviors of non-spherical particles
by using the variational argument and assumption of the marginal
stability. The variational argument gives a typical amplitude of the
minimal eigenvalue near the jamming transition point, while the marginal
stability requires that the minimal eigenvalue vanishes. For spherical
particles, this approach can reproduce the correct scaling of the excess
contact number and shear modulus~\cite{wyart2005effects}. Since our
argument is very similar to that of spherical particles, we first give a
summary of the variational argument of spherical particles.

For spherical particles, the interaction potential is given by the
$\Delta\to 0$ limit of Eqs.~(\ref{165606_24Jan19}).  We consider a
quadratic expansion of the potential around an equilibrium
position:
\begin{align}
\delta V_N = \sum_{ij=1}^N\delta\bx_i\pdiff{V_N}{\bx_i\partial \bx_j}\delta\bx_j
 = \sum_{ij=1}^N\delta\bx_i
\left[
 \pdiff{^2 v(h_{ij}(0))}{h_{ij}(0)^2}\pdiff{h_{ij}(0)}{\bx_i}\pdiff{h_{ij}(0)}{\bx_j}
 + \pdiff{v(h_{ij}(0))}{h_{ij}(0)}\pdiff{^2h_{ij}(0)}{\bx_i\partial \bx_j}
 \right]\delta\bx_j.
\end{align}
The second term in the square bracket is the so-called pre-stress, which
is proportional to the pressure and reduces the
eigenvalues~\cite{alexander1998}.  If we neglect the pre-stress, the
quadratic expansion of the potential can be rewritten as
\begin{align}
 \delta V_N = \frac{k}{2}\sum_{\alpha}^{Nz/2}\delta r_\alpha^2,
\end{align}
where $k$ denotes the characteristic stiffness of the potential, $\alpha
= (i,j)$ denotes the contact pair, $z$ denotes the contact number per
particle, and $\delta r_{ij} =
\hr_{ij}\cdot(\delta\bm{x}_i-\delta\bm{x}_j)$. We shall start from an 
isostatic configuration $z=2d$, where the system has a zero mode and
thus $\lambda_{\rm min}=0$. To obtain a configuration above the jamming
transition point, we add $N\delta z/2$ extra contacts. With this
setting, one can express the minimal eigenvalue in a variational form:
\begin{align}
 \lambda_{\rm min} \sim 
 k\min_{\delta\bx_i}
  \frac{\sum_{\alpha=1}^{Nd}\delta r_\alpha^2
 + \sum_{\alpha=Nd +1}^{N\delta z/2}\left(\delta r_\alpha - y_\alpha\right)^2}{\sum_{i=1}^N \delta\bx_i\cdot\delta \bx_i},
\end{align}
where $y_\alpha = \varepsilon$ denotes the rest length of the additional
contacts.  Combining an appropriate linear transformation and finite $N$
scaling, one can estimate the typical amplitude of $\lambda_{\rm min}$
for $\delta z\ll 1$ as~\cite{yan2016}
\begin{align} 
\lambda_{\rm min} \sim k\delta z^2.\label{115708_4Nov19}
\end{align}
So far, we have neglected the effects of the pre-stress.  For spherical
particles, one can show that the pre-stress always gives a negative
contribution to the eigenvalues~\cite{wyart2005effects}. As the
pre-stress is proportional to the pressure, $\partial v(h_{ij}(0))/\partial h_{ij}(0)\sim -p$,
we get
\begin{align}
 \lambda_{\rm min} \sim k\delta z^2 - cp,\label{112928_4Nov19}
\end{align}
where $c$ denotes a positive constant. At the jamming transition point,
$\delta z=0$ and $p=0$, meaning that the system is marginal stable
$\lambda_{\rm min}=0$. We assume that this marginal stability persists
even above the jamming transition point; we require $\lambda_{\rm
min}\sim 0$ or at most $\lambda_{\rm min}\sim p$.  From this condition
and Eq.~(\ref{112928_4Nov19}), we have~\cite{wyart2005effects}
\begin{align}
 \delta z \sim p^{1/2}.\label{144022_4Nov19}
\end{align}
This reproduces the results of numerical simulations in two and three
dimensions~\cite{ohern2003}.

\subsection{Variational argument for non-spherical particles}

Now we apply the variational argument to non-spherical particles. We
first discuss the stability of a system consisting of non-spherical
particles. At the jamming transition point $p=0$, $Q_N$ vanishes, and
the number of constraints given by $V_N$ is
\begin{align}
 N_{V}^c &= \frac{Nz}{2},
\end{align}
where $N$ denotes the number of particles, and $z$ denotes the number of
contacts per particle. For non-spherical particles, $N_V^c$ is smaller
than the number of degrees of freedom, meaning that there are
unconstrained modes, which we hereafter refer to as the \textit{zero
modes}. The number of the zero modes is
\begin{align}
 N_0 = N(d+d_{\rm rot})-N_{V}^{c} = Nd_{\rm rot}-\frac{N}{2}\delta z,
\end{align}
where $\delta z = z-2d$, and $d_{\rm rot}$ denotes the number of
rotational degrees of freedom per particle. For $p>0$, $Q_N$ has a
finite value that would stabilize some of the zero modes. $Q_N$ gives
$N_{Q}^c= Nd_{\rm rot}$ number of constraints, the typical
stiffness of which is $k_R = \partial_{\bu_i}\partial_{\bu_j}Q_N \sim
p\Delta$.  This indeed suffices to stabilize the
zero modes: 
\begin{align}
 N_Q^c- N_0 = \frac{N}{2}\delta z > 0.
\end{align}

When $\delta z\ll 1$, the system is considered to be nearly isostatic if
one only takes into account the zero modes. In this case, we can apply
the variational argument in Ref.~\cite{yan2016} to the zero modes, as in
the case of spherical particles near the jamming transition point. By
repeating a similar argument used in the previous sub-section, we obtain
\begin{align}
 \lambda_{\rm min} \sim k_{R}\delta z^2 \sim c_1 p\Delta \delta z^2,
\end{align}
where $c_1$ is a constant. Here we assume that $c_1>0$, which can be
validated by the explicit calculations for ellipsoids interacting with
the Gay-Berne potential, and breathing particles~\cite{BP2018}.
However, this assumption is not validated for some shapes of particles.
We shall discuss this point in Sec.~\ref{152804_8Jul19}. If we take
into account the second order term of $\Delta$, we have
\begin{align}
 \lambda_{\rm min} \sim p\left[c_1\Delta \delta z^2 + c_2 \Delta^2 +
 O(\Delta^3)\right],\label{145853_28Jun19}
 \end{align}
where all terms should be proportional to $p$, because the zero mode
vanishes as $p$. Although the first order term of $\Delta$ can be
positive, the pre-stress, in general, destabilizes the system due to the
structural buckling~\cite{alexander1998}. Therefore, it is natural to
assume that the second order term gives a negative contribution $c_2<0$.
The marginal stability requires $\lambda_{\rm min}\sim 0$, meaning that the
first and second terms in the RHS of Eq.~(\ref{145853_28Jun19}) should be canceled
each other, which leads to the scaling of $\delta z$:
\begin{align}
 \delta z \sim \Delta^{1/2}\sim (\A-1)^{1/4}.\label{140052_26Jan19}
\end{align}
\begin{figure}
\centering \includegraphics[width=.8\textwidth]{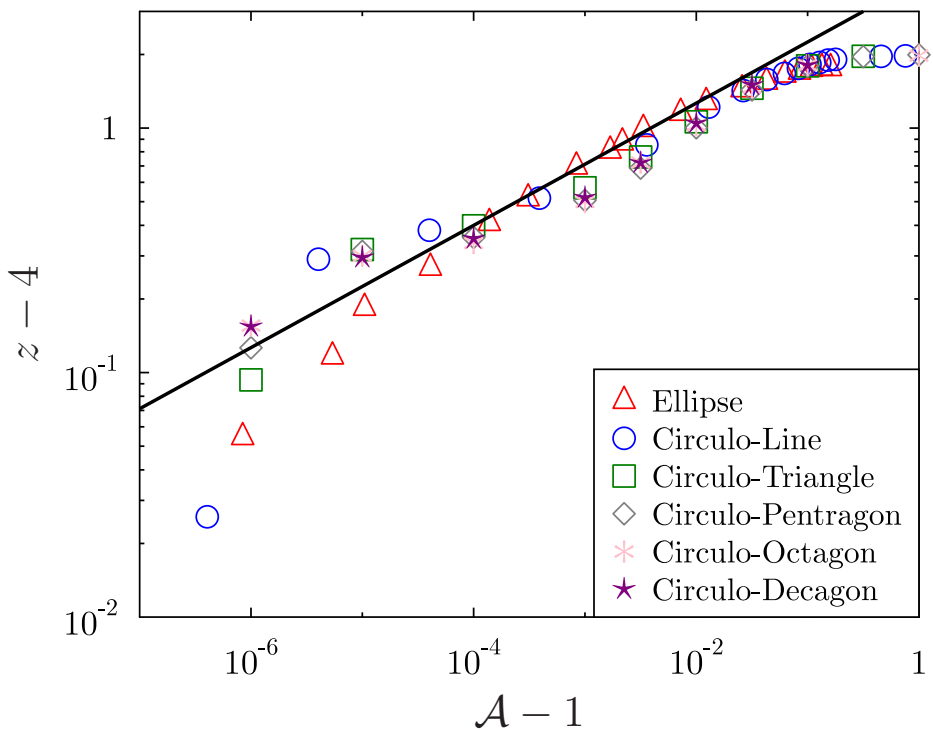}
\caption{Scaling of the contact number of non-spherical particles.  The
symbols denote the numerical results, and the solid line denotes the
theoretical prediction $z-4\sim (\A-1)^{1/4}$. Data for non-spherical
particles are taken from Ref.~\cite{vanderwerf2018}. }
\label{140029_26Jan19}
\end{figure}
In Fig.~\ref{140029_26Jan19}, we compare the theoretical prediction,
Eq.~(\ref{140052_26Jan19}), with the numerical results of various shapes
of non-spherical particles in Ref.~\cite{vanderwerf2018} where the jammed
configurations were generated by the binary search combining isotropic
compression and decompression. We get an excellent agreement with the
theory and numerical result, though there are visible deviations for
$\A-1\ll 1$, which might be originated from the lack of statistics,
numerical precision, or finite size effects.

In the $\Delta\to 0$ limit, Eq.~(\ref{140052_26Jan19}) should be
smoothly connected to the result of spherical particles
Eq.~(\ref{144022_4Nov19}).  From this condition, one can decide the
scaling form of $\delta z$ as
\begin{align}
\delta z = \Delta^{1/2}\mathcal{Z}(\Delta^{-1}p),\label{164137_1Feb19}
\end{align}
where
\begin{align}
 \mathcal{Z}(x) &=
 \begin{cases}
  const & (x\ll 1),\\
  x^{1/2} & (x\gg 1).
 \end{cases}
\end{align}
$\Z(x)$ is a finite and regular function at $x=0$, and thus one can
expand it as $\Z(x) = \Z(0) + \Z'(0)x +\cdots$, which leads to
\begin{align}
 z -z_J \sim \frac{p}{\Delta^{1/2}},
\end{align}
where $z_J = 2d + \Delta^{1/2}\Z(0)$. This is again consistent with a
numerical result of ellipsoids~\cite{schreck2010}.

We now turn our attention to the scaling of the shear modulus $G$. In
the standard numerical procedure to calculate $G$, one first imposes the
small strain and then minimizes the energy. Comparing the resultant
energy with that of the undeformed one, one can calculate
$G$~\cite{ohern2003}. Previous numerical and theoretical investigations
prove that the square root singularity $G\sim p^{1/2}$ appears near the
jamming transition of spherical
particles~\cite{ohern2003,wyart2005effects}. We here discuss how this
scaling is altered for non-spherical particles. We assume that the
imposed shear excites only the zero modes because the typical energy of
those modes is much smaller than the other near the jamming transition
point. When $\Delta\ll 1$, the zero modes mainly consists of the
rotational degrees of freedom.  The typical displacement $\delta u_i$
caused by the imposed shear strain $\delta\gamma$ is roughly $\delta u_i
\sim \delta\gamma/\Delta$. After the minimization, the energy difference
caused by the shear can be expressed as
\begin{align}
 \delta V_N \sim \delta Q_N \sim \min_{\bm{y}_i} \sum_{i=1}^N \sum_{\alpha=1}^{d_{\rm rot}}
 k_{i}^\alpha\left(\delta \tilde{u}_i^\alpha + y_i^\alpha\right)^2,\label{192356_26Jan19}
 \end{align}
where $k_i$ and $\delta\tilde{\bu}_i$ denotes the eigenvalue and
eigenvector of
$\partial_{\bu_i^\alpha}\partial_{\bu_i^\beta}g_i(\bu_i)$, respectively.
$\bm{y}_i=\{y_i^1,\cdots, y_i^{d_{\rm rot}}\}$ denotes the vector spanned
by the $N_0$ zero modes to be chosen to minimize the energy. Using the
standard technique of the linear algebra, one can eliminate $N_0$ terms
among $Nd_{\rm rot}$ terms in Eq.~(\ref{192356_26Jan19}) (see. Ch.7 in
Ref.~\cite{wyart2005rigidity}). Thus, the typical amplitude of $\delta
V_N$ is
\begin{align}
 \delta V_N \sim k_R (Nd_{\rm rot}-N_0)\left(\frac{\delta \gamma}{\Delta}\right)^2
 \sim  \frac{N p\delta z \delta\gamma^2}{\Delta}.
\end{align} 
The shear modulus $G$ is then calculated as 
\begin{align}
 G\sim \frac{\delta V_N}{N\delta \gamma^2} \sim \frac{p}{\Delta^{1/2}}.\label{203456_26Jan19}
\end{align}
The result is consistent with the numerical results of
ellipsoids~\cite{mailman2009}. In the $\Delta\to 0$ limit,
Eq.~(\ref{203456_26Jan19}) smoothly connects to the result of spherical
particles~\cite{ohern2003,wyart2005effects},
\begin{align}
 G\sim p^{1/2}.
\end{align}
This requires the following scaling form:
\begin{align}
 G = \Delta^{1/2}\mathcal{G}(\Delta^{-1}p),\label{011728_28Jan19}
\end{align}
where the scaling function $\G(x)$ satisfies
\begin{align}
 \G(x) &=
 \begin{cases}
  x & (x\ll 1),\\
  x^{1/2} & (x\gg 1).
 \end{cases}
\end{align}
\begin{figure}
\centering \includegraphics[width=.8\textwidth]{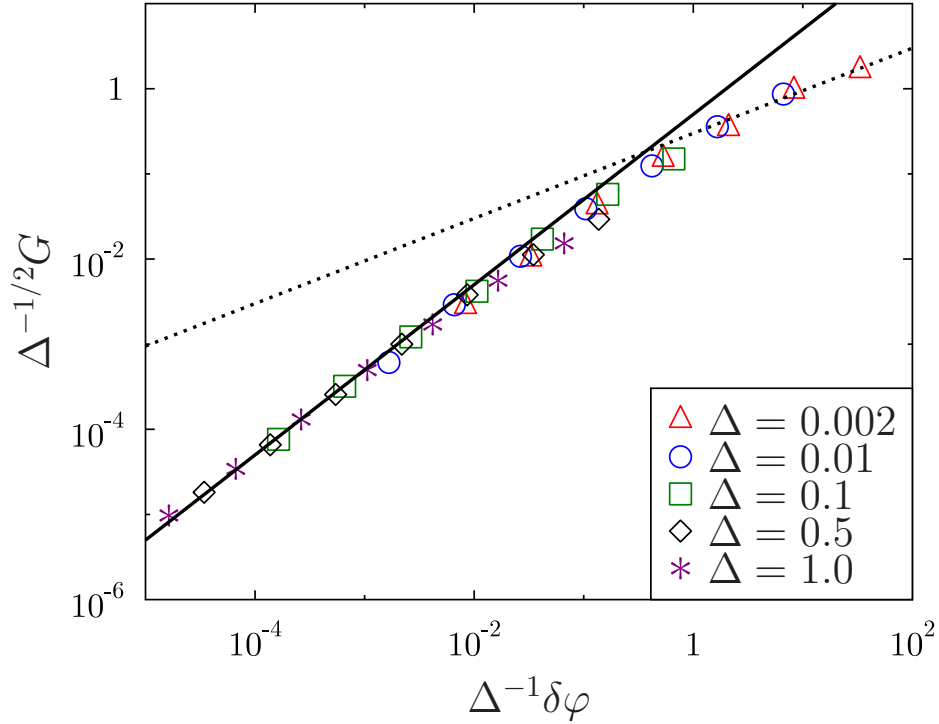} \caption{Saling
plot of the shear modulus of ellipsoids.  The symbols denote the
numerical results.  The solid and dotted lines denote the theoretical
predictions $G\sim \delta\varphi$ and $G\sim\delta\varphi^{1/2}$,
respectively. Data are taken from Ref.~\cite{mailman2009}.}
						       \label{011655_28Jan19}
\end{figure}
In Fig.~\ref{011655_28Jan19}, we confirm our scaling prediction for
ellipsoids interacting with harmonic potential, where $\Delta$ can be
identified with the aspect ratio, and $p\propto \delta\varphi\equiv
\varphi_J-\varphi$.  One can see that the data of different aspect
ratios are collapsed on a single curve, proving the validity of the
scaling prediction Eq.~(\ref{011728_28Jan19}).

\section{Connection between non-spherical particles and breathing particles}
\label{152720_8Jul19}

The above argument can be generally applied for models having extra
degrees of freedom, in addition to the translational degrees of freedom.
Besides non-spherical particles, another interesting model that belongs
to the same universality class is the so-called breathing particle (BP)
model~\cite{brito2018universality}. The model consists of $N$
polydisperse particles, the radii of which can change continuously. The
interaction potential of the model is given by
\begin{align}
 V_N= U_N + Q_N,\label{152005_2Feb19}
\end{align}
where
\begin{align}
 U_N &= \sum_{i<j}\frac{h_{ij}^2}{2}\theta(-h_{ij}),
 & h_{ij}&= r_{ij}-R_i-R_j,\label{152030_2Feb19}
\end{align}
and
\begin{align}
 Q_N &= \frac{k}{2}\sum_i \left(R_i-R_i^0\right)^2
 \left(\frac{R_i^{0}}{R_i}\right)^2.\label{152035_2Feb19}
\end{align}
We chose the stiffness $k$ so that the standard deviation of the radii
is proportional to $\Delta$:
\begin{align}
 \Delta \propto \sqrt{\frac{1}{NR_0^2}\sum_{i}\left(R_i-R_i^{0}\right)^2}.
\end{align}
From the saddle point condition, $\partial_{R_i}V_N=0$, one can infer
that
\begin{align}
 k \sim \frac{p}{\Delta}.
\end{align}
By introducing the new variable $u_i\equiv (R_i-R_i^0)/\Delta$,
Eqs.~(\ref{152030_2Feb19}) and (\ref{152035_2Feb19}) can be rewritten as
\begin{align}
U_N &= \sum_{i<j}\frac{h_{ij}^2}{2}\theta(-h_{ij}),& h_{ij}&=
r_{ij}-R_i^0-R_j^0 + \Delta(u_i+u_j),\new Q_N &= \frac{k_R}{2}\sum_i
u_i^2 \left(\frac{R_i^0}{R_i^0+\Delta u_i}\right)^2,\label{111309_2Aug19}
\end{align}
where
\begin{align}
 k_R = \Delta^2 k  \sim \Delta p.\label{115557_20Jun19}
\end{align}
Note that $k_R$ now has the same order as that of non-spherical
particles. Thus, one can repeat the same arguments in the previous
sections for non-spherical particles, which leads to the same critical
exponents~\cite{brito2018universality}. The numerical implementation of
the BP model is rather simpler, and the calculation time is shorter than
those of non-spherical particles, as the extra degrees of freedom are
simple scalar variables. For this reason, we shall use the BP in the
numerical experiments in the following sections, instead of non-spherical
particles. To obtain jammed configurations of the BP system, we use the
FIRE algorithm to find the inherent structures of the potential
Eq.(\ref{111309_2Aug19}). We use the Barendsen barostat to find them at
a fixed pressure. All the details are explained in 
Ref.~\cite{brito2018universality}.

\section{Universal scaling of the gap and force distributions near isostatic point}
\label{152733_8Jul19} Here we show that the gap and force distributions
exhibit the universal scaling behavior near the isostatic point.

\subsection{Definition of the distribution functions}
Here we investigate the gap distribution 
\begin{align}
 \rho(h) &\equiv \frac{1}{N}\ave{\sum_{i<j}\delta(h_{ij}-h)}.
\end{align}
At the zero temperature $T=0$, $\rho(h)$ has a gap at
$h=0$~\cite{ohern2003}. For this reason, it is convenient to define 
distributions for the positive and negative $h$, separately. We define
a positive gap distribution
\begin{align}
 g(h) &\equiv \theta(h)\frac{\rho(h)}{\int_0^\infty dh \rho(h)},\label{003451_22Jan19}
\end{align}
and a force (normalized negative gap) distribution
\begin{align}
 P(f) &\equiv \theta(-h)\frac{\rho(h)\diff{h}{f}}{\int_{-\infty}^0
 \rho(h)\diff{h}{f}df},\label{003504_22Jan19}
\end{align}
where $\theta(x)$ is the Heviside function, and $f= -h/p$. It is well
known that at the jamming transition point of spherical particles,
$g(h)$ and $P(f)$ exhibit the power laws for small $h$ and
$f$~\cite{donev2005,charbonneau2012uni}:
\begin{align}
 g(h) &\sim h^{-\gamma},\new
 P(f) &\sim f^{\theta}.\label{145304_28Jan19}
\end{align}
$\gamma$ does not depend on the spatial dimensions $d$ for $d\geq 2$ and
follows the mean-field prediction
$\gamma=0.41$~\cite{charbonneau2014fractal}, while $\theta$ exhibits the
weak $d$ dependence due to the localized
excitations~\cite{charbonneau2015jamming}.  Below, we discuss that the
power law is generally truncated at finite $h$ and $f$ if the system is
not isostatic.

\subsection{Finite size scaling}

Here we first describe the distribution functions of spherical particles
when the number of particles $N$ is finite.  Then, in the next
subsection, we show that the scaling of finite $N$ can be generalized to
the scaling of non-isostatic systems, including non-spherical particles
at the jamming transition point.

The minimal gap $h_{\rm min}$ of the $N$ particle system is calculated
by using the extreme statistics~\cite{kallus2016}
\begin{align} 
\int_0^{h_{\rm min}}g(h)dh \sim h_{\rm min}^{1-\gamma} \sim \frac{1}{N}
 \Rightarrow h_{\rm min} \sim N^{-\frac{1}{1-\gamma}}.\label{151001_28Jan19}
\end{align}
When $h\ll h_{\rm min}$, $g(h)$ quickly decreases, implying that the
power law divergence of $g(h)$, Eq.~(\ref{145304_28Jan19}), is truncated
at $h\sim h_{\rm min}$. Thus, the scaling form of $g(h)$ at finite $N$
would be~\cite{kallus2016}
\begin{align}
 g(h) &\sim \begin{cases}
	     N^{\mu\gamma} p_0^{+}(h N^{\mu}) & (h\ll  N^{-\mu}),\\
	     h^{-\gamma} & (h\gg N^{-\mu})
	    \end{cases},\label{165351_2Jun19}
\end{align}
where $p_0^{+}(x)$ is a regular and finite function, and
\begin{align}
 \mu = \frac{1}{1-\gamma}.
\end{align}
\begin{figure}
\centering \includegraphics[width=0.9\textwidth]{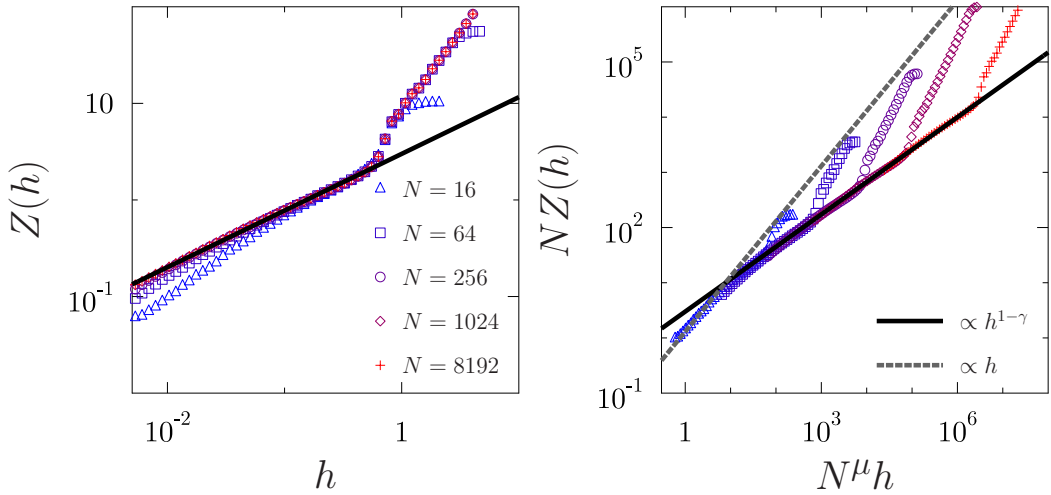}
  \caption{Cumulative gap distribution (left) and its sclaing plot
  (right).  Symbols are results of numerical simulations of the harmonic
  potential at the jamming point for different system sizes, and the
  lines are theoretical predictions. } \label{123653_3Jun19}
\end{figure}
We perform the numerical simulation for two-dimensional harmonic
spheres to test the above conjecture.
Instead of $g(h)$, we observe the
cumulative distribution $Z(h) = \int_0^{h'} dh' g(h')$ to improve the
statistics. From Eq.~(\ref{165351_2Jun19}), $Z(h)$ should satisfy
\begin{align}
 Z(h) &\sim \begin{cases}
	     N p_0^{+}(h N^{\mu}) & (h\ll N^{-\mu}),\\
	     h^{1-\gamma} & (h\gg N^{-\mu})
	    \end{cases}.\label{124624_3Jun19}
\end{align}
In Fig.~\ref{123653_3Jun19}, we plot $Z(h)$ and its scaling form for
several $N$.  The excellent scaling collapse justifies our scaling
argument.

\subsection{General scaling form of the distribution functions near isostatic point}

\begin{figure}
\centering \includegraphics[width=0.9\textwidth]{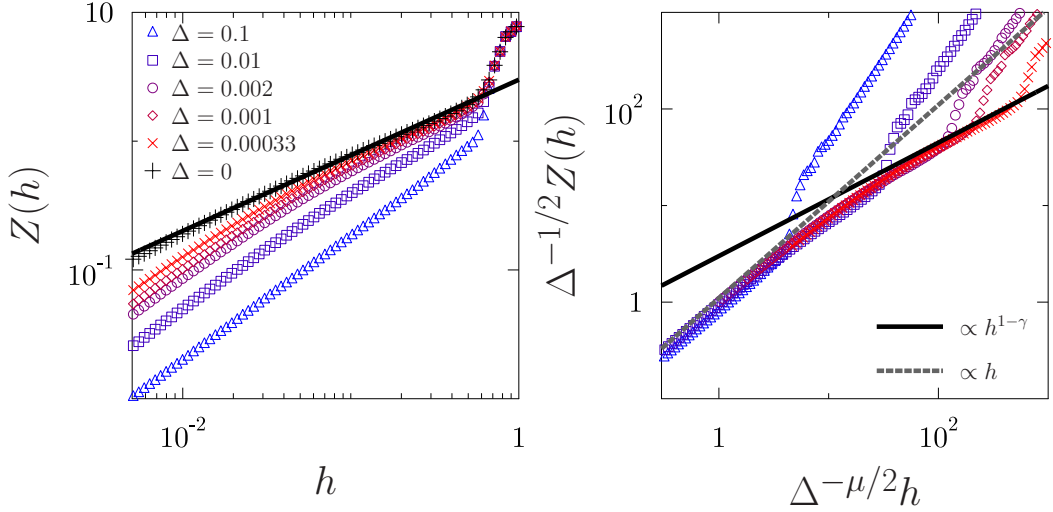}
\caption{Cumulative gap distribution (left) of non-spherical particles
and its sclaing plot (right).  Symbols are results of numerical
simulations of the BP system for different values of variance of the
 radii $\Delta$ and a system with N=8192 particles. Lines are
the theoretical predictions.}  \label{154547_5Jun19}
  \end{figure}

We want to generalize the above argument for more general systems close
to the isostatic point $\delta z\ll 1$. For this purpose, we
shall consider some function $F(h)$ which has the following scaling form
for $\delta z\ll 1$:
\begin{align}
F(h) = \delta z^\alpha \FF(\delta z^\beta h),\label{105738_6Aug19}
\end{align}
where $\alpha$ and $\beta$ denote the critical exponents we want to
determine from the finite size scaling. The extensive numerical
simulations of spherical particles prove that the scaling like
Eq.~(\ref{105738_6Aug19}) persists up to $\delta z =
1/N$~\cite{goodrich2012,goodrich2016}, suggesting that for a finite
size system at the jamming transition point, we have
\begin{align}
F(h) = N^{-\alpha} \FF(N^{-\beta} h).\label{110616_6Aug19}
\end{align}
In other words, the scaling for $\delta z\ll 1$ can be obtained by
substituting $N=\delta z^{-1}$ into the result of the finite size
scaling.
From Eqs.~(\ref{165351_2Jun19}) and
(\ref{124624_3Jun19}), we have
\begin{align}
 g(h) &\sim \begin{cases}
	     \delta z^{-\mu\gamma} p_0^{+}(h\delta z^{-\mu}) & (h\ll \delta z^{\mu})\\
	     h^{-\gamma} & (h\gg \delta z^{\mu})
	    \end{cases},\label{111321_2Jul19}
\end{align}
and 
\begin{align}
 Z(h) &\sim \begin{cases}
	     \delta z^{-1} p_0^{+}(h \delta z^{-\mu}) & (h\ll \delta z^{\mu})\\
	     h^{1-\gamma} & (h\gg \delta z^{\mu})
	    \end{cases}.\label{111325_2Jul19}
\end{align}
We propose that the above equations hold for any system sufficiently
near the isostatic point, \textit{i.e.}, $\delta z\ll 1$.  We shall test
this conjecture for the BP at the jamming transition point, at which we
have shown that $\delta z \sim
\Delta^{1/2}$~\cite{brito2018universality}. In Fig.~\ref{154547_5Jun19},
we show $Z(h)$ and its scaling form of the BP at the jamming transition
point. The excellent scaling collapse justifies the validity of
Eqs.~(\ref{111321_2Jul19}) and (\ref{111325_2Jul19}). Note that the same
equation of Eq.~(\ref{111321_2Jul19}) holds exactly in the case of
the mean-field model of non-spherical
particles~\cite{brito2018universality,ikeda2019mean} and the spherical
particles slightly above the jamming transition point, where $\delta z
\sim p^{1/2}$~\cite{franz2017}.

For the force distribution $P(f)$, one can apply a similar argument, leading to 
\begin{align}
P(f) &\sim \begin{cases} \delta z^{\theta\nu}p_0^{-}(f\delta z^{-\nu}) &
	    (f\ll \delta z^{\nu})\\ f^\theta & (f\gg \delta z^{\nu})
	   \end{cases},
\end{align}
where $p_0^{-}(x)$ denotes a finite and regular function, and we have
introduced the critical exponent by
\begin{align}
\nu &= \frac{1}{1+\theta}.
\end{align}
The numerical justification of $P(f)$ is rather tricky because one
should carefully separate the localized and extended modes to compare them
with the theoretical prediction~\cite{charbonneau2015jamming}, which we
leave for future work.

\section{Vibrational density of states}
\label{163348_3Jul19}

The vibrational density of states $D(\omega)$ characterizes the low
temperature physics of solids~\cite{kittel1996}. In this section, we
investigate $D(\omega)$ of the BP model near the jamming transition
point, which exhibits the same scaling as non-spherical particles.  The
results shown in the following are made for a system with $N=484$
particles. 

\subsection{Characteristic frequencies}
We first describe the qualitative shape of $D(\omega)$ and define the
characteristic frequencies.
\begin{figure}
\centering \includegraphics[width=1.0\textwidth]{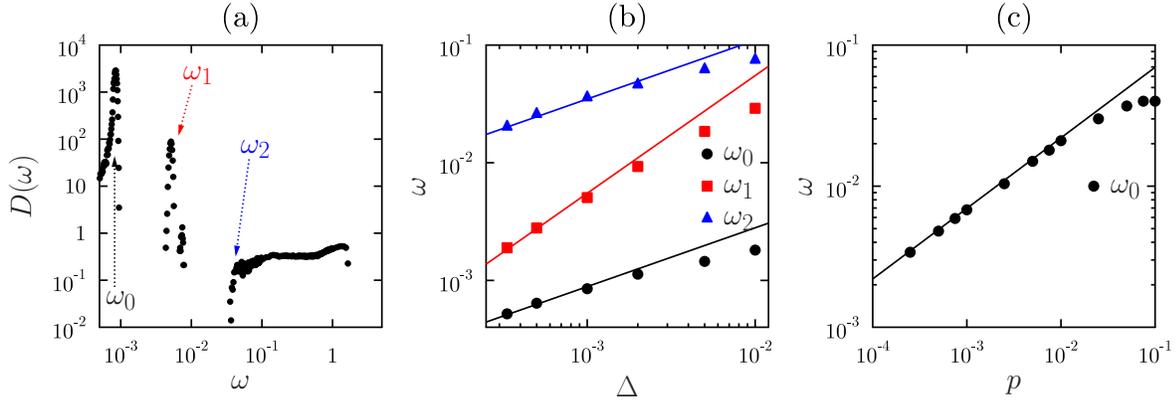} \caption{ (a)
Density of states $D(\omega)$ for the BP system at $\Delta=10^{-3}$ and
$p=10^{-4}$ and the definition of the characteristic frequencies.  (b)
The $\Delta$ dependence of the characteristic frequencies. Lines
are the theoretical predictions, $\omega_0 \propto \Delta ^{1/2}$, $\omega_1
\propto \Delta $, and $\omega_2 \propto \Delta ^{1/2}$ (c) The $p$
dependence of the characteristic frequencies.  Line is the theoretical
prediction, $\omega_0 \propto p^{1/2}$.  } \label{151912_18Jun19}
\end{figure}
In Fig.~\ref{151912_18Jun19} (a), we show the typical behavior of
$D(\omega)$ of the BP model. $D(\omega)$ consists of the three separate
parts: (i) the lowest band at $\omega_0$, (ii) the intermediate band at
$\omega_1$, and (iii) the highest band starts from $\omega_2$. In
Fig~\ref{151912_18Jun19} (b), we show the $\Delta$ dependence of the
characteristic frequencies, $\omega_0$, $\omega_1$, and $\omega_2$.  The
characteristic frequencies are well fitted by the following power laws
(see solid lines):
\begin{align}
 &\omega_0 \sim \Delta^{1/2},&
 &\omega_1 \sim \Delta, &
 &\omega_2 \sim \Delta^{1/2}.
\end{align}
In Fig.~\ref{151912_18Jun19} (c), we show the $p$ dependence of
$\omega_0$.  We found that
\begin{align}
\omega_0\sim p^{1/2},
\end{align}
while $\omega_1$ and $\omega_2$ remain constant (not shown). The above
scaling is the same as ellipsoids if we identify
$\Delta$ with the aspect ratio~\cite{schreck2012}, which is another
evidence that the BP and ellipsoids belong to the same
universality class.

\begin{figure}
  \centering \includegraphics[width=0.8\textwidth]{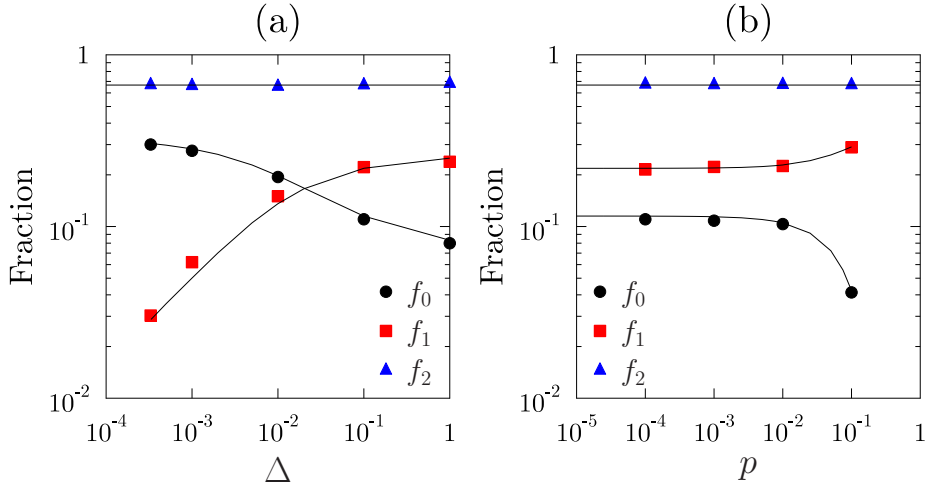} \caption{The
 weights of the three bands of the $D(\omega)$ as the example shown in
 Fig.~(\ref{151912_18Jun19})-a. Symbols denote the numerical results for
 the BP system, and the solid lines denote the theoretical predictions,
 see main text.  (a) The $p$ dependence at fixed variance
 $\Delta=10^{-1}$.  (b) The $\Delta$ dependence at fixed pressure $p =
 10^{-4}$.}  \label{204902_18Jun19}
\end{figure}

Using the previous theoretical
analysis~\cite{brito2018universality,ikeda2019mean}, we can understand
the above scalings in the three regions. (i) The lowest band corresponds
to the zero modes stabilized by the positive part of the pre-stress. As
the pre-stress scales as $k_R \sim p\Delta$, Eq.~(\ref{115557_20Jun19}),
the characteristic frequency of the mode is $\omega_0\sim \sqrt{k_R}\sim
p^{1/2}\Delta^{1/2}$. (ii) The intermediate band corresponds to the
breathing motion of the BP or the rotation of ellipsoids.  Therefore,
the characteristic frequency is $\omega_1\sim \sqrt{\partial_R^2
V_N}\sim \Delta$. (iii) The highest band corresponds to the
translational degrees of freedom. As in the case of spherical particles,
the characteristic frequency is proportional to $\delta
z$~\cite{wyart2005rigidity}. Using Eq.~(\ref{140052_26Jan19}), we get
$\omega_2\sim \Delta^{1/2}$. Herewith we recover the above numerical
results. To give further evidence to support the above picture, we
calculate the weights of each band by numerically integrating
$D(\omega)$. If the above description is correct, one should have the
following equations:
\begin{align}
&N_0 = N \left(1-\frac{\delta z}{2}\right),&
&N_1 = \frac{N\delta z}{2},&
&N_2 =  dN,\label{121418_20Jun19}
\end{align}
where $N_i$ denotes the number of the modes included in the $i$-th band.
Since the total number of modes is $3N$, the fraction $f_i$ of modes in
each band is given by $f_i=N_i/3N$.  In Fig.~\ref{204902_18Jun19}, we
show the $\Delta$ and $p$ dependencies of $f_i$ and compare with the
theoretical prediction Eqs.~(\ref{121418_20Jun19}). We obtain quite good
agreement.

\section{Summary and discussions}
\label{152804_8Jul19}

\begin{figure}
\centering \includegraphics[width=.8\textwidth]{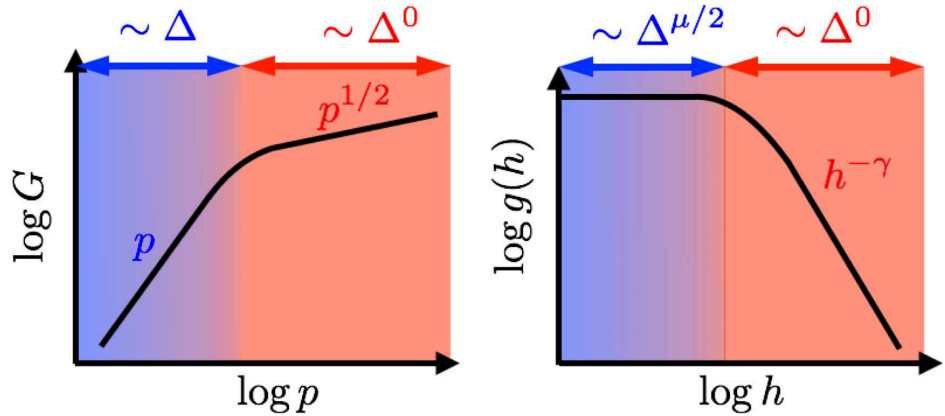} \caption{ Summary
 of the scaling prediction.  $\Delta$ denotes the linear deviation from
 the spherical particles.  (left) The shear modulus $G$ as a function of
 the pressure $p$. $G$ exhibits the linear scaling $G\sim p$ for $p\ll
 \Delta$, while $G\sim p^{1/2}$ for $p\gg \Delta$. (right) The gap
 distribution $g(h)$. $g(h)$ exhibits the power law $g(h)\sim
 h^{-\gamma}$ for $h\gg \Delta^{\mu/2}$, while it converges to a finite
 value for $h\ll \Delta^{\mu/2}$.} \label{140122_2Aug19}
\end{figure}

In this paper, we investigated the jamming transition of non-spherical
particles and breathing particles.  Using both numerical and scaling
arguments, we confirmed that the critical behavior of the jamming of
non-spherical particles and breathing particles is qualitatively
different from that of spherical particles.  In the left panel of
Fig.~\ref{140122_2Aug19}, we summarize our scaling prediction for the
shear modulus $G$.  Note that, for non-spherical particles ($\Delta>0$),
$G$ always shows the linear pressure $p$ dependence sufficiently near
the jamming transition point ($p\ll \Delta$), while it exhibits the
square root dependence for spherical particles ($\Delta=0$). This means
that the critical exponent of $G$ changes discontinuously at $\Delta=0$
from one to one half; in other words, the small asphericity is enough to
change the universality class of the jamming transition.  We also show
that non-spherical particles and breathing particles are not critical at
the jamming transition point in terms of the gap distribution $g(h)$,
see the right panel of Fig.~\ref{140122_2Aug19}. The power law
divergence of $g(h)$ is truncated at finite $h$, and thus the gap
distribution is finite and analytic even at the jamming transition
point. This is a sharp contrast to spherical particles, where the power
law divergence of $g(h)$ persists up to $h=0$. Furthermore, we fully
characterized the scaling of the characteristic frequencies of the
density of states near the jamming transition point, which are again
dramatically different from those of spherical particles.

There are still several open questions. A tentative list is the
following:
\begin{itemize}
 \item It is important to understand the
       rheological properties of the system near the jamming transition
       point. It has been shown that the divergence of the viscosity is
       strongly connected to the lowest excitation of the density of
       states~\cite{lerner2012unified}. As discussed in
       Sec. \ref{163348_3Jul19}, the density of states of non-spherical
       particles is very different from that of spherical particles,
       which would change the critical exponent of the viscosity compared to
       that of spherical particles. A further study of this point is
       left as an open problem.

 \item In this work, we assumed that the pre-stress gives a positive
       contribution for the first order of $\Delta$, $c_1>0$. This
       assumption is violated for non-spherical particles consisting
       of spherical particles, such as dimers. In this case, the
       rotational motions of non-spherical particles can be identified
       with the translational motions of particles that consists
       non-spherical particles. Therefor, the system becomes isostatic
       and exhibits the same scaling of that of spherical
       particles~\cite{schreck2010,papa2013}. Furthermore, Platonic
       solids are also known to be isostatic at the jamming transition
       point~\cite{smith2011}. Further studies are necessary to uncover
       when the assumption $c_1>0$ can be validated.

 \item In this work, we assumed that the two particles could have at
       most one contact. This assumption is correct for particles of
       convex shape. However, for particles of non-convex shape, the two
       particles can have more than two contacts. The extension of our
       work for such non-convex shape particles is an interesting future
       work. We believe that the study along this direction would be a
       promising way to introduce the effect of the friction, which is
       considered to be originated from the surface roughness of the
       constituent particles.
\end{itemize}

\ack We thank F.~Zamponi and P.~Urbani for previous joint
research~\cite{brito2018universality} on which this work is based.

This project has received funding from the European Research Council
(ERC) under the European Union's Horizon 2020 research and innovation
programme (grant agreement n.723955-GlassUniversality).  This work was
supported by a grant from the Simons Foundation (\#454953, Matthieu
Wyart and \#454955).

\section*{References}

 \bibliographystyle{iopart-num.bst}
 \bibliography{reference}

\end{document}